\PassOptionsToPackage{table}{xcolor} 
\documentclass[twocolumn]{aastex631}

\usepackage{amsmath}

\makeatletter
\@ifpackageloaded{multirow}{}{%
  \usepackage{multirow}
}
\makeatother

\begin{document}

\title{Wind Accretion in Massive Binaries Experiencing High Mass Loss Rates: I. Dependency on Mass Ratio and Orbital Period}

\shorttitle{}
\shortauthors{B. Mukhija and A. Kashi}

\author[0009-0007-1450-6490]{Bhawna Mukhija}
\affiliation{Department of Physics, Ariel University, Ariel, 4070000, Israel}
\email{bhawnam@ariel.ac.il}

\author[0000-0002-7840-0181]{Amit Kashi}
\email{kashi@ariel.ac.il}
\affiliation{Department of Physics, Ariel University, Ariel, 4070000, Israel}
\affiliation{Astrophysics, Geophysics, and Space Science (AGASS) Center, Ariel University, Ariel, 4070000, Israel}

\begin{abstract}

We run numerical simulations to study high-power wind accretion in a massive binary system during a high mass loss event.
The system consists of an evolved primary star with a zero age main sequence mass of $ M_{1} = \rm 100~M_{\odot}$ and a hot secondary star with a mass ranging from $ M_{2} = \rm 30-80~M_{\odot}$, orbiting in a circular orbits with periods between 455 and 1155 days. We initiate a weak eruption event with mass loss at a rate of $10^{-3}~\rm {M_{\odot}}\rm~yr^{-1}$ for 1.5 years. During this event, a fraction of the mass lost by the primary is accreted onto the secondary, with the accretion rate being dependent on the orbital and stellar parameters. From the set of simulations, we derive an analytical relation describing the dependence of the mass accretion rate on the orbital period and stellar mass ratio. We also identify the transitional orbital period for which Roche lobe overflow begins to dominate over wind accretion. We find that accretion leads to a reduction in the effective temperature of the secondary star. However, the mass  average accretion rate we obtain in the simulations is low enough for the secondary to remain in thermal equilibrium and avoid radial expansion.
\end{abstract}

\keywords{stars: massive --- stars: mass loss --- stars: winds, outflows--- stars:binaries---stars; accretion---method: numerical}
    
\section{Introduction}

Massive stars are frequently born in binary and higher multiplicity systems \citep[e.g.,][]{1998NewA....3..443V,2009AJ....137.3358M, 2013ARA&A..51..269D, 2017A&A...598A..84A, 2024ARA&A..62...21M}. It suggested that up to approximately 70 \%  of O-type stars may undergo mass exchange or merge with a companion before reaching the end of their evolution \citep{2022MNRAS.516.3297S}. The interactions between components in binary or multiple star systems, leading to mass and angular momentum transfer, play a role in determining the fundamental parameters and outcomes of both stars \citep[e.g.,][]{2012Sci...337..444S, 2014ApJS..215...15S, 2017ApJS..230...15M}. In older open clusters, most binaries with a Blue Straggler Star (BSS) \citep[e.g.,][]{2000ASPC..198..517M, 2009Natur.462.1032M, 2014ApJ...783L...8G, 2015ApJ...814..163G} and a Blue Lurkers (BL) \citep[e.g.,][]{2019ApJ...881...47L, 2021ApJ...908..229L} component tend to have wide orbits with periods $ \approx 1000$ days, too far apart for Roche lobe overflow (RLOF) to occur. In such systems, the mass transfer (MT) process occurs through the wind accretion(e.g., Bondi-Hoyle-Lyttleton (BHL) wind accretion \citep[e.g,][]{1944MNRAS.104..273B, 1952MNRAS.112..195B, 2004NewAR..48..843E}).

Wind accretion is the dominat accretion process in the evolution of detached binary systems, such as symbiotic stars \citep[e.g.,][]{2016A&A...588A..83S, 2021MNRAS.501..201H, 2023A&A...680A..60S, 2025ApJ...980..224V, 2025arXiv250211325M}, precursor peculiar red giants \citep[e.g.,][]{escorza2018binaryinteractionredgiant, Escorza_2020}, Algol-type stars, and massive X-ray binaries \citep[e.g.,][]{2015A&A...579A.111K, El_Mellah_2019, Hainich_2020, 2025A&A...695A.117N}. Recently, \citet{2025A&A...695A.117N} analyzed 21 WR+O binary systems to assess whether MT commenced during the main sequence (Case A) or in a later evolutionary stage (Case B). By estimating the MT efficiency ($\beta$) and angular momentum loss, the study offers new insights into their evolutionary pathways leading to the formation of X-ray binaries and double black hole systems.

\citet{2017MNRAS.464..775K} used 3D hydrodynamic simulations to study the wind accretion process during the periastron passage of $\eta$ Carina. The simulations modeled the interactions between the stellar winds of the two stars, focusing on the accretion of the primary's wind onto the secondary. The study found that dense gas filaments began to accrete onto the secondary star, and that this accretion leads to significant changes in the secondary's wind, resulting in observable spectroscopic variations. The estimated mass accreted onto the secondary was approximately $ 10^{-6}~\rm M_{\odot}$ during each periastron passage. In \citet{2020MNRAS.492.5261K}, they simulated the wind collision and accretion processes in the massive binary system HD 166734 using 3D hydrodynamic models. This system consists of two blue supergiants, orbiting each other with a period of about 34.5 days and an eccentricity of 0.618. The simulations revealed that near periastron passage, the primary wind suppresses the secondary's wind, leading to accretion onto the secondary. Later on, \citet{2022MNRAS.516.3193K} performed numerical simulations of wind accretion in a massive colliding wind binary system, consisting of a luminous blue variable (LBV) primary and a Wolf-Rayet (WR) secondary. By varying the LBV's mass-loss rate, they explored different wind momentum ratios ($\eta$) and track accretion onto the secondary. For $0.001 \lesssim \eta \lesssim 0.05$, they found sub-BHL accretion, with rates following a power law in the static case. When $\eta \lesssim 0.001$, accretion becomes continuous and reaches BHL levels.

Recently \citet{2024ApJ...966..103P} performed 3D simulations of subsonic accretion onto a small, absorbing object moving through a uniform medium. For an adiabatic index $\gamma = 5/3$, the accretion rate is independent of Mach number and depends only on the object's mass and gas entropy. The results match BHL predictions, confirming isentropic, shock-free flow.

\begin{table*}
    \centering
    \begin{tabular}{|l|c c c c c c| c|}
    \hline
    \hline
    & \multicolumn{6}{c|}{Companion stars} & Primary star \\  
    \hline
    
    Stellar parameter &  30$\rm M_{\odot}$ & 40$\rm M_{\odot}$ & 50$\rm M_{\odot}$ & 60$\rm M_{\odot}$ & 70$\rm M_{\odot}$ & 80$\rm M_{\odot}$ & 100$\rm M_{\odot}$  \\
    \hline
    
     $  M ~( \rm M_{\odot}$) & 29.07 & 37.06 & 43.74 & 49.37 & 54.44 & 58.45 & 63.29    \\ 
     $\log T_{\rm eff}~ (\rm K)$ & 4.57 & 4.58 & 4.55 & 4.50 & 4.42 & 4.40 & 4.28  \\
     $\log R ~(\rm R_{\odot}) $ & 0.98 & 1.12 & 1.26 & 1.46 & 1.67 &1.78 & 2.07  \\
     $\log g ~(\rm cm~s^{-2})$ & 3.93 & 3.77 & 3.55 & 3.22 & 2.82 & 2.63 & 2.08  \\
     $\log L ~(\rm L_{\odot})$ & 5.20 & 5.50 & 5.70 & 5.86 & 6.00 & 6.10 &  6.25  \\
     $\log \dot{M}~(\rm {M_{\odot}}\rm~yr^{-1})$ &-6.39 & -5.87 & -5.56 &  -5.43 & -5.39 & -5.11 & -4.12  \\
    \hline 
    \hline
    \end{tabular}
    \caption{Stellar parameters correspond to the $ \rm 30-80~M_{\odot}$ companion stars, and $ \rm 100~M_{\odot}$ primary star and at the initial stage of wind accretion. The rows from first to last represent: initial mass of the star ($ M$), surface temperature ($ T_{\rm eff}$), surface gravity ($ g $), surface luminosity ($ L$), and mass loss via stellar winds ($ \dot{M}$), respectively.}
    \label{T1}
\end{table*}

In \citet{2024ApJ...974..124M}, we investigated the impact of a Giant Eruption (GE) on a massive star by artificially simulating the eruption and examining how the star's stellar properties and evolutionary track on the HR diagram change during the eruption phase. We applied a high mass loss rate of approximately $\rm 0.15~M_{\odot}~yr^{-1}$ to the star’s outer layers over 20 years. The results indicated that the star’s luminosity decreased and its temperature increased due to the mass loss process. It focused on the mass-losing star in the binary system. Later on in \citet{2025ApJ...986..188M}, we performed numerical simulations that were
run to analyze how the companion star reacted to the accreted material and how its structure and evolution were altered due to the accretion process. A grid of massive accreting stars ranging from
20 to 60 $\rm M_{\odot}$ was set with accretion rates varying from $10^{-4}$ to $10^{-1}~\rm M_{\odot}~yr^{-1}$ over a duration of 20 years. We simulated a single star and assumed that it is a companion star in a binary system, accreting material at a very high rate due to the GE of the primary star by the wind accretion process.

In this paper, instead of modeling the stars individually, we model them as a binary system. We evolve a binary system where the primary star has a strong wind and the companion star accretes some of the material through the wind accretion. We quantify the amount of mass accreted onto the secondary star due to this process. We then run more simulations varying the orbital period and mass ratio to examine how the accretion rate depends on these parameters. To conduct these numerical simulations, we use public available stellar evolution code \textsc{mesa} ( Modules for Experiments in Stellar Astrophysics (\textsc{mesa} version- r.23.05.1)) \citep[][]{2011ApJS..192....3P, 2013ApJS..208....4P, 2015ApJS..220...15P, 2018ApJS..234...34P, 2019ApJS..243...10P, 2023ApJS..265...15J}. 

The paper is organized as follows.
Section \ref{2} provides background information of wind accretion, and \ref{3} represents the physical ingredients of our modeling.  Section \ref{4} presents the results of our analysis. Finally, Sections \ref{5}, \ref{6} provide our discussion and a summary of the findings.

\section{Wind Accretion}
\label{2}

The wind accretion rate is utilized in a wide range of scenarios, including wind MT in binary systems \citep{1988A&A...205..155B}, gas accretion by stars in stellar clusters \citep[e.g.,][]{2002A&A...383..491T, 2007ApJ...661..779L}, and even in galactic contexts \citep{2000MNRAS.318.1164S}. It explains the massive object's motion relative to a uniform fluid, which gravitationally focuses and accretes material onto itself from the surroundings \citep[e.g.,][]{1939Natur.144.1019H, 1944MNRAS.104..273B, 1952MNRAS.112..195B, 2004NewAR..48..843E}. In numerous systems, accretion rates derived from the BHL prescription are consistent with observations, for e.g., a direct BHL estimate well reproduces the mean X-ray luminosity of the wind-fed HMXB Vela X-1 \citep[e.g.,][]{2018A&A...610A..60S, 2021A&A...652A..95K}; similarly, BHL-like efficiencies inferred from 3D wind-accretion models match observations in symbiotic binaries \citep{2017MNRAS.468.3408D}.

Therefore, the wind accretion process plays a vital role in understanding the dynamics and evolution of binary star systems where MT occurs through stellar winds. If the primary star loses mass at a certain rate via its stellar wind, the secondary star in the system can potentially accrete some of this material as it orbits through the wind.

The wind velocity $v_{w}$, and density $\rho_{w}$ of primary star are
 \begin{equation}
    v^{2}_{\rm w} \approx 2 \beta_{\rm w} \frac{GM_{1}}{R_{1}} ,~~~~ \\
    \rho_{\rm w} = \frac{\dot M_{\rm w}}{4\pi a^{2}v_{\rm w}},
\label{BHL-rho}
 \end{equation}

where $\beta_{\rm w}$ represents the wind momentum flux parameter (proportional coefficient to the escape velocity from the surface), which is dependent on the spectral type of the star \citep{2001A&A...369..574V}.
$M_{1}$ is the mass of the primary star, $R_{1}$ is the stellar radius of the primary star, G is the gravitational constant, $\dot M_{w}$ is the mass loss rate (wind rate) of the primary star and $a$ is the binary separation. The orbital velocity of the binary system is given by
\begin{equation}
    v^{2}_{\rm orb} = \frac{G(M_{1} + M_{2})}{a} ,
    \label{BHL-v}
\end{equation}
where $M_{2}$ is the mass of the secondary star.The accretion rate onto the companion determined via the wind accretion is given by
\begin{equation}
     \dot M_{\rm acc} = 4\pi \rho_{\rm w} \frac{G^{2}M^{2}_{2}}{(v^{2}_{\rm s} + v^{2}_{\rm w} )^{3/2}},
     \label{BHL-Macc}
 \end{equation}
where $v_{s}$ is the speed of sound in the cloud of gas.

\section{The physical ingredients of numerical Simulation}
\label{3}

We use 6 binary systems with an orbital period varying from $P$ = 455  to 1155 days, having the primary stellar mass $ M_{1} =\rm 100~M_{\odot}$ and companions stellar masses uniformly ranging from $M_{2} = 30- \rm 80~M_{\odot}$. Firstly, we evolve the primary and companion star as a single star, till their post-MS phase.  We choose the phase just after the MS, which is shown by the dot marks on all evolutionary tracks in Figure \ref{fig_1} and use this point profile for the binary evolutions.
At this stage, both the primary and the companion stars have the same age. However, since the primary has a higher initial mass than the companion, its surface temperature is lower at this point in the evolution. Therefore, in our binary system, the primary star can be considered the cooler component, while the companion star is the hotter component. During the single star evolution, the star evolve with a lower rotational velocity of $ 0.1\Omega_{\rm crit}$ at solar metallicity $Z$=0.0142 \citep{2009ARA&A..47..481A}. To simulate wind accretion accretion in a binary system, we first evolved single-star models until the end of the MS phase. The stellar structures at that stage were then used as input for the binary evolution. From this point, we implemented wind accretion by incorporating the wind accretion formalism directly into the binary calculation. Since wind accretion is not part of the default \textsc{mesa} binary, we introduced additional prescriptions to activate mass accretion through the BHL equations \ref{3}. Table \ref{T2} contains the parameters for wind accretion during the binary evolution. 

We simulated the binary evolution over $t$ = 1.5 years, where the primary star loses mass at a constant rate of $\dot{M}_{1} = \rm 10^{-3}~M_{\odot}~yr^{-1}$, with a wind velocity of $v_{1\,w} = \rm 507~km~s^{-1}$.
The mass loss rate and wind velocity are in the ranges of the primary star of $\eta$~Car \citep[e.g.,][]{2017MNRAS.464..775K}.
This is an artificial mass loss rate that we implemented in \textsc{mesa}. During this simulation, we do not include any wind mass-loss prescriptions for either star, as our primary objective is to study accretion onto the secondary star via the wind accretion mechanism, driven solely by the primary’s artificially imposed mass loss. The expected wind-driven mass loss of the primary at this evolutionary stage is much smaller. Including the wind mass loss of the secondary would introduce additional complexity into the system, as the companion, being hotter, would likely have a stronger intrinsic wind, which could reduce the efficiency of wind accretion \textit{, bringing it to the sub-BHL accretion regime \citep{2022MNRAS.516.3193K}}. Since our focus is on studying the effect of the imposed primary outflow, we defer such effects to future studies. We aim to investigate how the amount of accreted material affects the secondary star. To isolate the effects of rotationally enhanced winds, we consider only non-rotating models during our binary evolution \citep[e.g.,][]{2000A&A...361..159M, 2012ARA&A..50..107L, 2014A&A...564A..57M}. 

We use the OPAL type II opacity tables \citep{1996ApJ...464..943I}, which allow for time-dependent variation in the metal abundance in our model. An additional wind hook is applied to model stellar winds. The time-dependent convection  (\textsc{tdc}) theory is utilized to treat convection, using a fixed convective mixing length parameter of $\alpha_{\mathrm{MLT}} = 1.5$ \citep[e.g.,][]{2018ApJS..234...34P, 2023ApJS..265...15J}. This formalism is also applied to the sub-surface convective regions that develop near opacity bumps. We activate the \textsc{mlt++} prescription, which modifies the convective treatment by reducing the superadiabatic temperature gradient in regions where standard \textsc{mlt} would otherwise predict inefficient convection, allowing for a more stable and physically realistic treatment in radiation-dominated envelopes. In our stellar models, convective boundaries are determined using the Ledoux criterion, which incorporates the stabilizing effect of composition gradients by comparing the radiative and adiabatic temperature gradients. \textsc{mesa} consider semi-convection as a diffusive process, with the diffusion coefficient based on the formulation of \citet{1983A&A...126..207L}. We adopt a constant semi-convective efficiency parameter of $\alpha_{\rm sc} = 1$, following the approaches of \citet{2006A&A...460..199Y} and \citet{2019A&A...625A.132S}. Overshooting formalism is employed to address overshooting in the hydrogen-burning zone, and exponential overshooting is implemented above any convective core to prevent numerical instabilities \citep{2000A&A...360..952H}. The overshooting parameters are $f_{0}(1) = 0.01$, and $f_{1}(1) = 0.345$ respectively.

\begin{table*}[ht]
\vspace*{\fill} 
\centering
\caption{Binary stellar parameters adopted for wind accretion modeling.}
\label{T2}
\begin{tabular}{|l|l|c|c|}
\hline
\hline
Symbol & Parameter& Value & Units \\
\hline
\( a \) & Orbital period & 455 – 1155 & days \\
\( e \) & Orbital eccentricity & 0 & – \\
\( \dot{M}_\mathrm{donor} \) & Mass-loss rate from donor star & \(10^{-3}\) & \(\rm  M_\odot~\mathrm{yr}^{-1} \) \\
\( \alpha \) & Wind accretion efficiency parameter & 1.5 & – \\
\( \beta \) & Wind velocity exponent & 1.25 & – \\
\( \eta_{\mathrm{max}} \) & Maximum wind transfer efficiency & 0.5 & – \\
\( f_{\mathrm{pextra}} \) & Pextra factor  & 1.5 & – \\
\( q \) & Mass ratio \( q = M_2 / M_1 \) & 0.3 – 0.8 & – \\
\hline
\hline
\end{tabular}
\vspace*{\fill} 
\end{table*}

\section{Results}
\label{4}
\begin{figure*}
  \includegraphics[trim= 2.0cm 0.0cm 3cm 1.0cm,clip=true,width=1\textwidth]{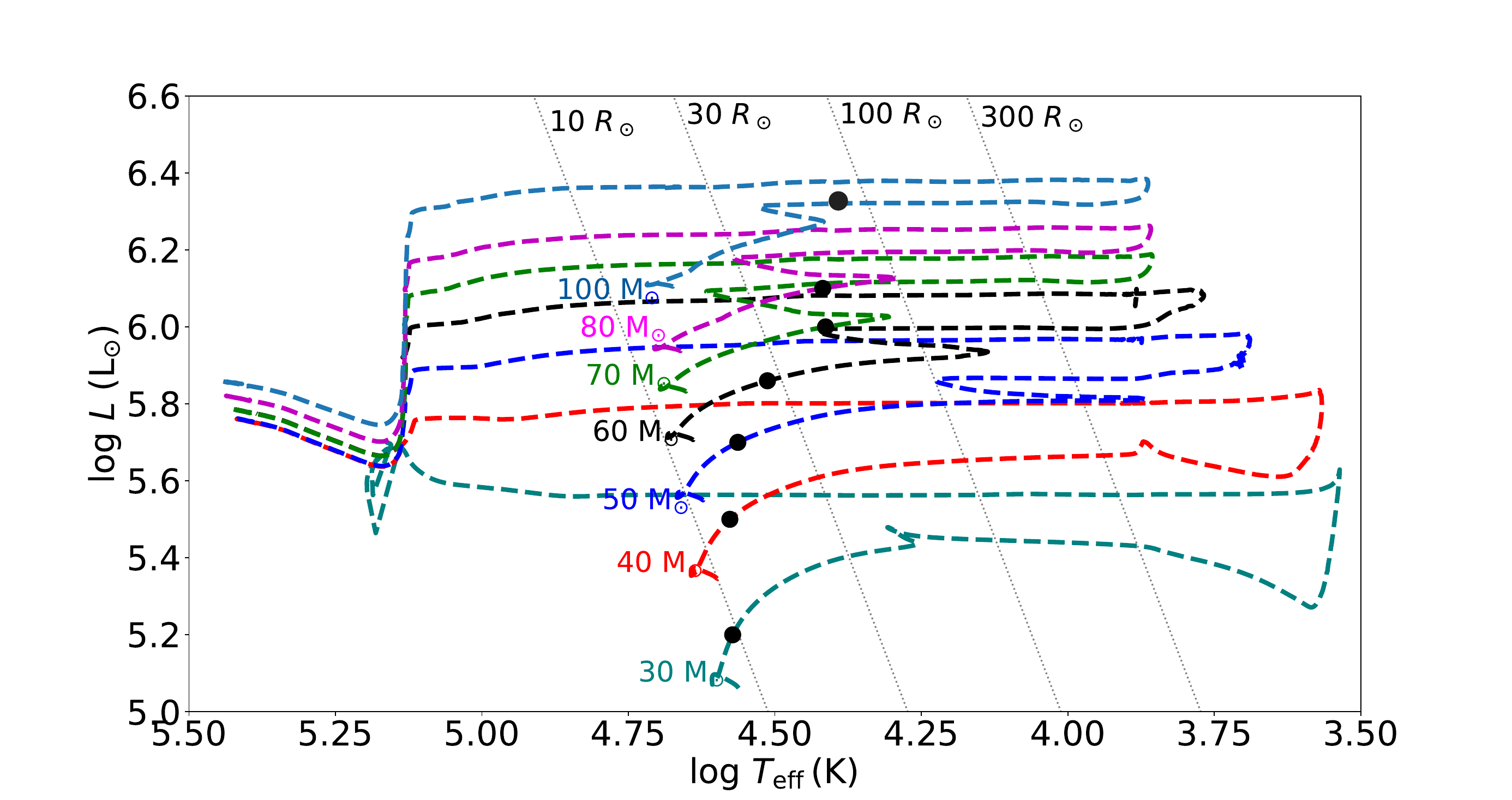} 
  \centering
 
\caption{Evolution tracks of the companion stars (ranging from 30 to 80 $\rm M_{\odot}$) and the primary star (100 $\rm M_{\odot}$) from ZAMS to the WR phase at Galactic metallicity by using \textsc{mlt++} formalism. The black dots on each evolutionary track indicate the star at the time we start the wind accretion simulation. The stellar parameters corresponding to those points are listed in Table~\ref{T1}.
}
 \label{fig_1}
\end{figure*}

We evolve binary systems using \textsc{MESA} to initiate wind accretion in circular orbits. The initial conditions, and location for the primary and companion stars undergoing wind accretion are provided in Table~\ref{T1}, and are also indicated as dotted points in Figure~\ref{fig_1}. The binary system parameters relevant for wind accretion are summarised in Table~\ref{T2}. Simulations are carried out for a duration of $t$ = 1.5 years, and the orbital period spans a range of periods from $P$ = 455 to 1155 days. For each value of the mass ratio $q = M_{2}/M_{1}$, ranging from 0.3 to 0.8 at ZAMS, we construct a comprehensive grid of models by varying the orbital period from 455 to 1155 days. This corresponds to six distinct systems simulated for each $q$ value. With six different mass ratios considered, the total number of simulations amounts to 42. This allows us to systematically analyze the mechanics of wind accretion and its impact on both the primary and companion stars, as well as its influence on binary parameters such as the orbital period and the mass ratio. From this grid, we select a representative system with a $M_{1}=100~\rm M_{\odot}$ primary and a $M_{2}=30~\rm M_{\odot}$ companion, and an orbital period of $P=555$ days, to study the wind accretion in detail in this Section below.

\subsection{Response of primary and companion stars during wind accretion}

In our simulation the primary star (i.e., $M_{1}$ ($M_{\rm ZAMS})= \rm 100~M_{\odot}$, and $M_{\rm wind accretion} = 63.29~\rm M_{\odot}$ at the onset the wind accretion)
undergoes mass loss at a constant rate $\rm 10^{-3}~M_{\odot}~yr^{-1}$ over $t$ = 1.5 years, for the wind accretion. The total mass lost from the primary star is 0.0015 $\rm M_{\odot}$ during this phase. The luminosity of the primary drops dramatically by an order of magnitude as it rapidly loses mass. The luminosity drop is not due to the decrease in the nuclear energy production (which stays roughly constant and even increases towards the end of the MT phase) but rather because most of the energy is trapped in the outer layers, which expand in response to rapid mass loss.

\begin{figure*}
  \includegraphics[trim= 0.0cm 0.0cm 0cm 0.0cm,clip=true,width=1\textwidth]{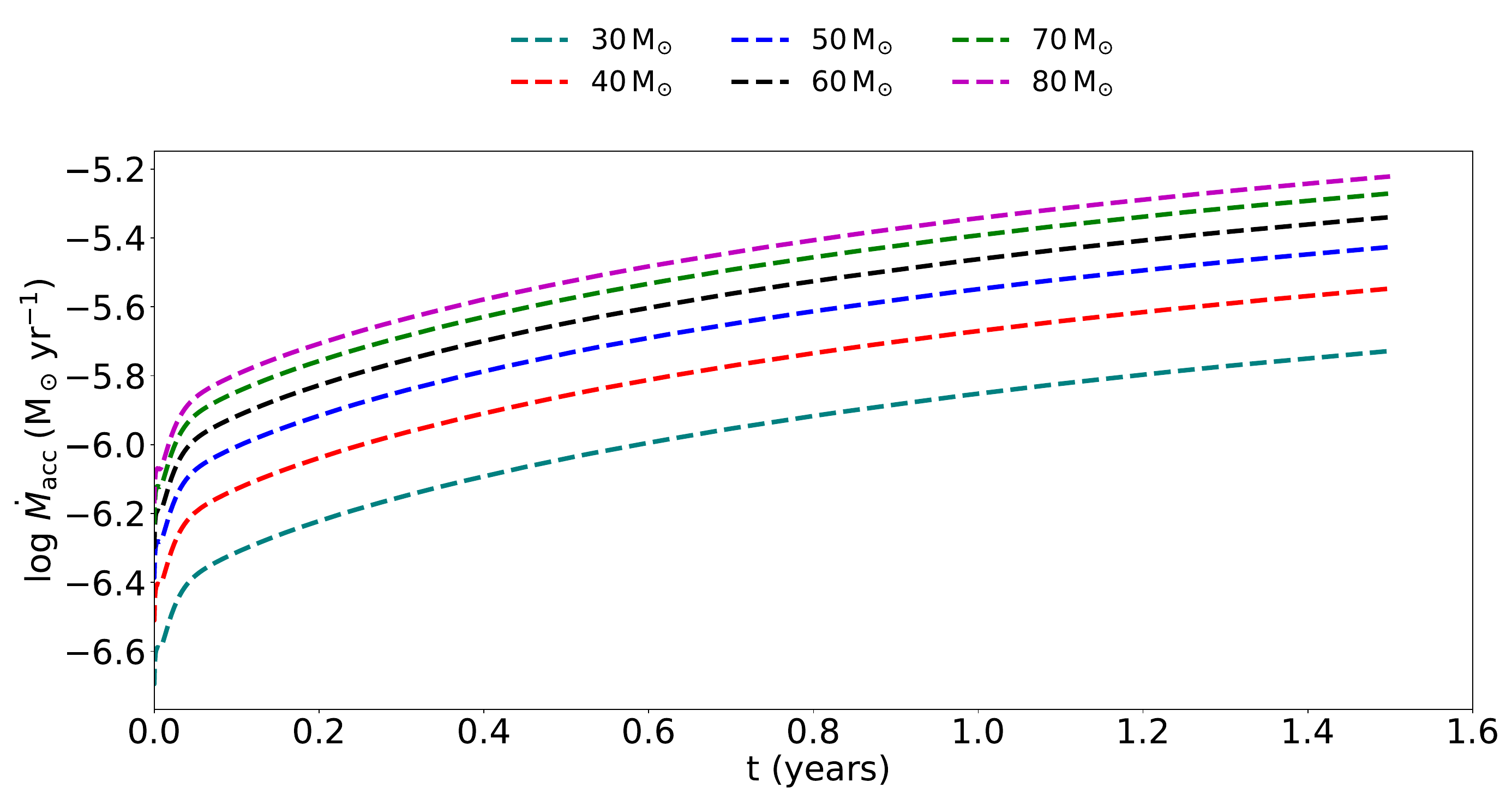} 
  \centering
  
\caption{average accretion rates for the different companion stars with an orbital period of $P=555$ days. The average accretion rates vary among companions due to the mass difference and gradually increase over the 1.5-year simulation period, as a result of the increased orbital velocity of the primary.}

\label{fig_2}
\end{figure*}

The companion star (i.e., $M_{\rm ZAMS} = \rm 30~M_{\odot}$, and $M_{\rm wind accretion}= \rm 29.07~M_{\odot} $ at onset of wind accretion) accreted mass from the primary star, via the wind accretion. The total mass gain on to the outer layer of the companion star is $ 1.72 \times 10^{-6}~\rm M_{\odot}$ during this phase over $t$ = 1.5 years. Its luminosity increases by a 0.392 \%, the radius expands by 0.508 \%, and the temperature increases by 0.323 \% .  The average accretion rate onto the companion over the 1.5 years is shown in Figure~\ref{fig_2} by the dashed cyan line. We note in Figure~\ref{fig_2} that the accretion rate is not strictly constant; instead, it exhibits a gradual increase throughout the simulation duration.  The mass-transfer efficiency is approximately 0.115 \%. Such a low efficiency indicates that the majority of the expelled material likely escaped the system. These results are consistent with expectations for wind-driven MT in wide binaries, where accretion is typically inefficient \citep[e.g.,][]{2004A&A...419..335N}.

It is important to notice that in our simulations, the final mass ratio at the onset of wind accretion, denoted as \( q_f \), can be empirically related to the initial mass ratio at ZAMS, \( q \), by the following relation:
\begin{equation}
q_f = q^{0.60 \pm 0.03}.
\end{equation}
This power law approximation provides a practical conversion for applying our figures and fitting relations, originally expressed in terms of \( q \), to the actual mass ratio values at the start of the wind accretion. 

During wind accretion, the primary star continuously loses mass, which results in a loss of gravitational energy and a subsequent decrease in its luminosity. Meanwhile, the companion star accretes some of the ejected material, gaining gravitational energy that is converted into thermal energy, thereby increasing its luminosity. The accretion process also affects the stellar structure, including changes in the radii and surface temperatures of the companion stars. For lower-mass companions (i.e., 30, 40, or 50~$\rm M_\odot$), the surface temperature remains nearly constant after accretion. In contrast, more massive companions exhibit a noticeable decrease in surface temperature, attributed to the expansion of their outer layers during the accretion phase. Additionally, as mass is transferred, the orbital velocity of the primary increases, leading to a variable accretion rate onto the companion as shown in Figure~\ref{fig_2}. However, the orbital velocity of the companion decreases, resulting in a gradual increase in the overall orbital period of the binary system.

As illustrated in Figure~\ref{fig_2}, increasing the companion mass from \(30\) to \(80\,\mathrm{M}_\odot\) (i.e., increasing \(q\)) leads to higher average accretion rates, despite a constant primary mass-loss rate and accretion duration. These results are consistent with those of \citet{2024ARA&A..62...21M}, who find that systems with more massive companions exhibit enhanced accretion efficiency and higher luminosity due to thermal disequilibrium. Similar trends have been reported in earlier studies by \citet{2004NewAR..48..843E, El_Mellah_2019}, further supporting the dependence of accretion efficiency on the companion’s mass.
 
\subsection{wind accretion rate and loss of angular momentum}\label{4.2}

\begin{figure*}
  \includegraphics[trim= 0.0cm 0.0cm 0cm 0.0cm,clip=true,width=1\textwidth]{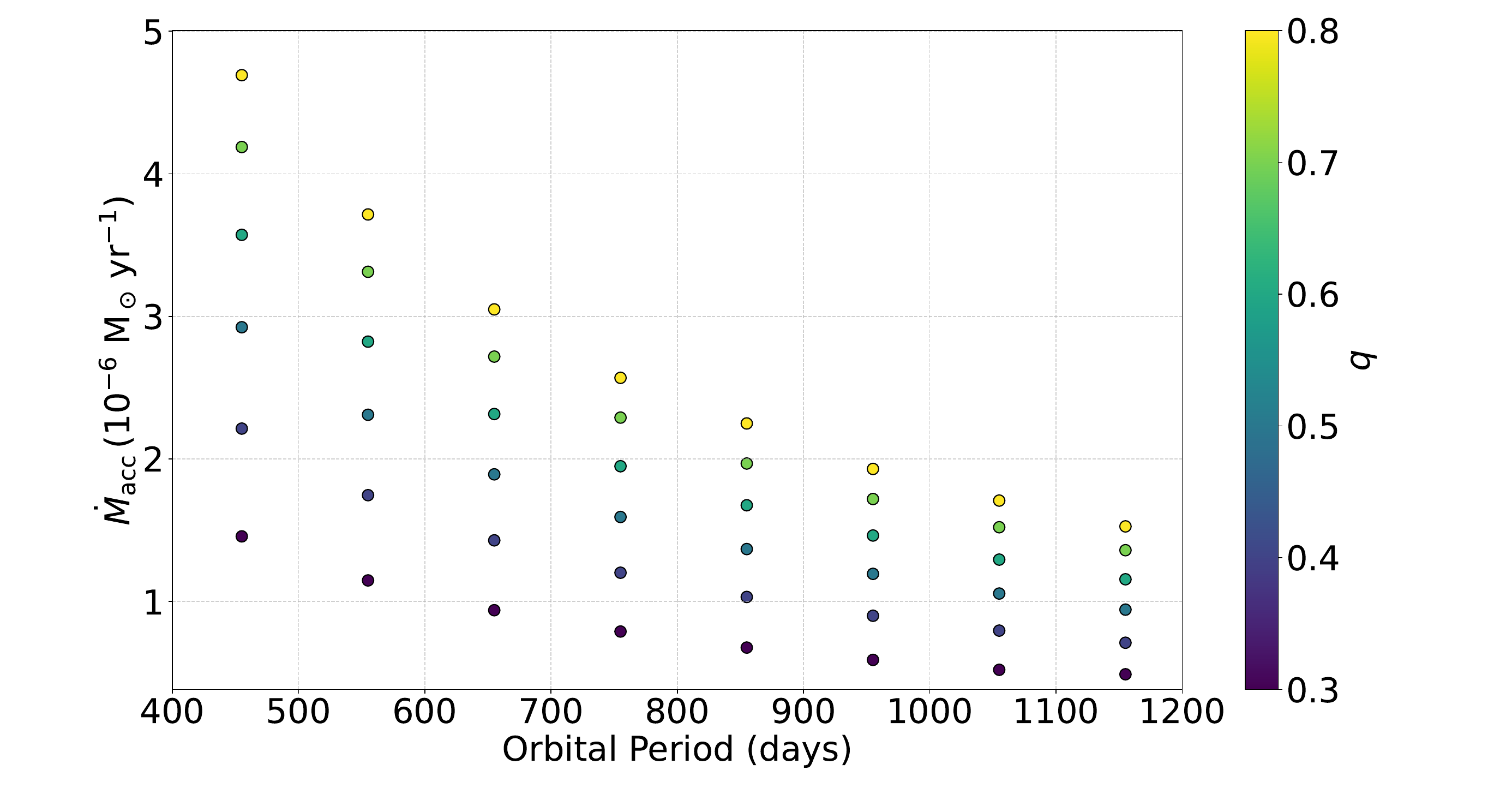} 
  \centering
  \caption{Comparison of average accretion rates and the luminosity–radius relationship for the full grid of six binary systems. It illustrates how the average accretion rate varies with both the orbital period and the mass ratio \( q \), showing that higher \( q \) values generally lead to increased accretion.}
\label{fig_3}
\end{figure*}

The primary star loses mass, and the companion star starts to gain mass throughout 1.5 years through wind accretion. We calculate the analytical average accretion rate during this time 

\begin{equation}
    \dot{M}_{\rm acc} = \frac{\Delta M}{\Delta t}
\end{equation}
Here $\Delta M$ is the mass accreted by the companion star during this phase, and $\Delta t$ is the period of this phase. Since the total accreted mass is $ 1.72 \times 10^{-6}~\rm M_{\odot}$ over 1.5 years of time period, this implies the $\dot M_{\rm acc}$ = $ 1.148 \times 10^{-6}~\rm M_{\odot}~yr^{-1}$ as mentioned in Table \ref{T4}  for the 30 $\rm M_{\odot}$ companion star.

For our \( q = 0.3 \) binary system, the wind accretion  yields an average accretion rate of approximately \( \dot{M}_{\rm acc} = 1.72 \times 10^{-6}~\rm M_\odot~\text{yr}^{-1} \) on the companion star. For the same system, the characteristic luminosity-driven mass-loss rate can be estimated by
\begin{equation}
\begin{split}
    \dot{M}_{\rm KH} &\approx 1.67 \times 10^{-3}
    \left(\frac{R_2}{9.54\, \rm{R_\odot}}\right)
    \left(\frac{L_2}{1.58\, \times 10^{5} \rm{L_\odot}}\right) \\
    & \times 
    \left(\frac{M_2}{29.07\,\rm{M_\odot}}\right)^{-1}
    ~\rm M_\odot~\text{yr}^{-1}.   
\end{split}
\end{equation}
Since  \( \dot{M}_{\rm acc} \ll \dot{M}_{\rm KH} \), the star is able to adjust thermally and remain in a stable state throughout the wind accretion (as mentioned in, e.g., \citealt[][]{2024MNRAS.531L..45Z}). We observe this consistent behavior across all of our binary models.

We simulated the same system (100 -- 30 $\rm M_{\odot}$) while varying the orbital period from $P=455$ to $P=1155$ days. Our results show that as the orbital period increases, the average accretion rate decreases as mentioned in Table \ref{T4} for \( q \) = 0.3. In contrast, a shorter orbital period leads to a higher average accretion rate due to the system being more compact, enhancing the accretion rate.
Additionally, we also find out that, at an orbital period of $P \simeq 355$ days, the system also undergoes RLOF MT. However, our focus is not on MT via RLOF. We identify the transition point at $P \simeq $ 355 days, where the star approaches the RLOF as shown in Table \ref{T3}.
At $P$ = 356 days, however, MT occurs solely through the wind accretion. Since our primary star is losing mass, there is also the loss of angular momentum from the system, which has a consequence on the orbital elements of the system itself. An isotropic mass loss from the mass-losing star leads to an increase in the semi-major axis. This was taken into account in the model for formation of Barium stars by \citet{1988A&A...205..155B}. It was also shown to be consistent with the observation through a statistical analysis of spectroscopic binaries containing red giants. In our simulations, there is a loss of angular momentum from the primary star, which causes the binary system to become wider after the wind accretion.

\subsection{A grid of the binary systems}

As illustrated in Figure~\ref{fig_2}, the average accretion rate shows dependence on the mass ratio \( q \), increasing progressively as \( q \) becomes larger. To account for this, we present all six binary systems in Figure \ref{fig_3} and derive a power-law relation that describes the dependence of average accretion rates on both orbital period and \( q \). Our findings indicate that as the \( q \) increases, the average accretion rate also increases. This is because a higher \( q \) value corresponds to a more massive companion star, which enhances the accretion process. Thus, the average accretion rate
\( \dot{M}_{\rm acc} \) in \( \rm M_\odot\, \text{yr}^{-1} \), is related to the orbital period $P$ (given in days), mass ratio \( q = M_2/M_1 \) and time period t (in years) by the power law
\begin{equation}
\dot{M}_{\rm acc}(t,P,q)
= A\left(\frac{P}{\mathrm{day}}\right)^{\alpha} q^{\beta} \left(\frac{t}{\mathrm{yr}}\right)^{\gamma},
\end{equation}
where \(A = 1.336\times10^{-2}\ {\rm M_\odot\,yr^{-1}}\), \(\alpha = -1.253\), \(\beta = 1.111\), and \(\gamma = 0.204\).

In our \(100\) – \(30\,\rm M_\odot\) binary system, we investigated how the onset of RLOF depends on the orbital period. By varying the period, we identified a clear transition: at \(P \lesssim 355\) days, both wind accretion and RLOF contribute to MT, while at \(P \gtrsim 356\) days, only wind accretion is present, as shown in Table \ref{T3}. We define this boundary as the transition point where RLOF becomes dynamically significant. Although wind accretion continues to operate during RLOF, its influence is secondary. 
Extending this analysis to systems with different mass ratios \(q = M_2/M_1\), we found that increasing \(q\) systematically shifts the transition point to shorter orbital periods. To characterize this behavior, we derived an empirical fitting formula for the critical period at which RLOF begins, based on our model data:

\begin{table}
\centering
\caption{Accretion mode as a function of orbital period in the simulated system for \(q = 0.3\).}
\begin{tabular}{|c|c|c|}
\hline
 $P$ (days) & Accretion Mode & Criterion \\
\hline
$< 355$ &  wind accretion + RLOF & $R_1 \geq R_{\rm L1}$ \\
$\sim 355$ & Transition Regime & $0.92 \lesssim R_1 / R_{\rm L1} < 1$ \\

$1155 - 356$ &  wind accretion & $R_1 / R_{\rm L1} < 1$ \\

\hline
\end{tabular}
\label{T3}
\end{table}

\begin{equation}
P_{\mathrm{RLOF}}^{\mathrm{fit}} \approx (108 \pm 1.2 )\cdot f(q)^{-3/2} \cdot \left( \frac{q}{1 + q} \right)^{1/2},
\end{equation}

\noindent where \( f(q) \) is the volume-equivalent Roche-lobe radius approximation given by \citet{1983ApJ...268..368E}:

\begin{equation}
f(q) = \frac{0.49\, q^{2/3}}{0.6\, q^{2/3} + \ln(1 + q^{1/3})}.
\end{equation}

This empirical relation captures the decreasing trend of the transition period with increasing mass ratio and provides a practical diagnostic for RLOF onset across different systems. Notably, the behavior exhibited by this fit shows strong qualitative agreement with the semi-analytic treatment presented by \citet{2024ARA&A..62...21M}, who find that Galactic metallicity massive stars can initiate RLOF at periods as short as \(\sim 1\) day on the ZAMS and as long as \(\sim 10\) years at maximum radial expansion. In our simulations, the transition to RLOF consistently occurs between \(300\) and \(400\) days, suggesting that our systems probe an intermediate evolutionary regime where envelope expansion is sufficient to bring the donor into contact with its Roche lobe. 


Figure~\ref{fig_5} presents a 3D representation of the wind accretion, showing the relationship between the average accretion rate, orbital period, and mass ratio \( q \) over a fixed simulation time of 1.5 years. The figure illustrates that higher values of \( q \) correspond to higher average accretion rates, while longer orbital periods are associated with reduced average accretion rates. This comprehensive view highlights the combined influence of binary parameters on the efficiency of wind accretion.

\begin{deluxetable*}{cccccccc}
\tablecaption{Summary of binary accretion parameters for different mass ratios \( q \) and orbital periods \( P \). The table lists the average accretion rate \( \dot{M}_{\rm acc} \) (in units of \( 10^{-6}~\rm M_\odot~\text{yr}^{-1} \)), binary separation (in \( \rm R_\odot \)), stellar radius of donor star \( R_1 \), Roche-lobe radius of donor star \( RL_1 \), and the Roche-lobe filling factor \( f = R_1 / RL_1 \).}
\label{T4}
\tablehead{
\colhead{\(q\)} & \colhead{\(q_f\)} & \colhead{\(P\) (days)} & \colhead{\(\dot{M}_{\rm acc} (10^{-6}~\rm {M_{\odot}}\rm~yr^{-1}) \)} & \colhead{Separation (\(R_\odot\))} & \colhead{\(R_1\) (\(R_\odot\))} & \colhead{\(RL_1\) (\(R_\odot\))} & \colhead{\(f = R_1/RL_1\)}
}
\startdata
\multirow{8}{*}{0.3} & \multirow{8}{*}{0.45} & 455 & 1.457 & 1125.34 & 9.59 & 503.65 & 0.019 \\
                     &                       & 555 & 1.148 & 1284.70 & 9.60 & 574.98 & 0.017 \\
                     &                       & 655 & 0.939 & 1434.73 & 9.61 & 642.12 & 0.015 \\
                     &                       & 755 & 0.789 & 1577.27 & 9.61 & 705.91 & 0.014 \\
                     &                       & 855 & 0.677 & 1713.64 & 9.62 & 766.95 & 0.013 \\
                     &                       & 955 & 0.590 & 1844.78 & 9.62 & 825.64 & 0.012 \\
                     &                       &1055 & 0.521 & 1971.41 & 9.62 & 882.31 & 0.011 \\
                     &                       &1155 & 0.489 & 2039.34 & 9.63 & 912.72 & 0.011 \\
\hline
\multirow{8}{*}{0.4} & \multirow{8}{*}{0.58} & 455 & 2.213 & 1156.87 & 13.27 & 492.78 & 0.027 \\
                     &                       & 555 & 1.746 & 1320.71 & 13.30 & 562.57 & 0.024 \\
                     &                       & 655 & 1.429 & 1474.94 & 13.32 & 628.26 & 0.021 \\
                     &                       & 755 & 1.202 & 1621.47 & 13.34 & 690.68 & 0.019 \\
                     &                       & 855 & 1.032 & 1761.66 & 13.35 & 750.40 & 0.018 \\
                     &                       & 955 & 0.900 & 1896.48 & 13.36 & 807.27 & 0.017 \\
                     &                       &1055 & 0.796 & 2026.66 & 13.37 & 863.27 & 0.015 \\
                     &                       &1155 & 0.711 & 252.78  & 13.37 & 917.00 & 0.015 \\
\hline
\multirow{8}{*}{0.5} & \multirow{8}{*}{0.69} & 455 & 2.924 & 1181.99 & 19.41 & 486.12 & 0.040 \\
                     &                       & 555 & 2.310 & 1349.38 & 19.45 & 554.97 & 0.035 \\
                     &                       & 655 & 1.892 & 1506.96 & 19.49 & 619.77 & 0.031 \\
                     &                       & 755 & 1.593 & 1656.68 & 19.51 & 681.35 & 0.029 \\
                     &                       & 855 & 1.368 & 1799.91 & 19.53 & 740.26 & 0.026 \\
                     &                       & 955 & 1.194 & 1937.65 & 19.55 & 796.91 & 0.025 \\
                     &                       &1055 & 1.056 & 2070.66 & 19.56 & 851.61 & 0.023 \\
                     &                       &1155 & 0.943 & 2199.52 & 19.56 & 904.61 & 0.022 \\
\hline
\multirow{8}{*}{0.6} & \multirow{8}{*}{0.78} & 455 & 3.572 & 1202.37 & 32.79 & 481.67 & 0.068 \\
                     &                       & 555 & 2.823 & 1372.65 & 32.84 & 549.88 & 0.060 \\
                     &                       & 655 & 2.315 & 1532.94 & 32.83 & 614.09 & 0.053 \\
                     &                       & 755 & 1.949 & 1685.24 & 32.92 & 675.10 & 0.049 \\
                     &                       & 855 & 1.675 & 1830.94 & 32.94 & 733.47 & 0.045 \\
                     &                       & 955 & 1.463 & 1971.06 & 32.95 & 789.60 & 0.042 \\
                     &                       &1055 & 1.294 & 2106.36 & 32.95 & 843.80 & 0.039 \\
                     &                       &1155 & 1.156 & 2237.44 & 32.96 & 896.32 & 0.037 \\
\hline
\multirow{8}{*}{0.7} & \multirow{8}{*}{0.86} & 455 & 4.188 & 1220.15 & 61.12 & 478.33 & 0.128 \\
                     &                       & 555 & 3.313 & 1392.95 & 61.38 & 546.07 & 0.112 \\
                     &                       & 655 & 2.718 & 1555.61 & 61.58 & 609.84 & 0.101 \\
                     &                       & 755 & 2.290 & 1710.16 & 61.75 & 670.42 & 0.092 \\
                     &                       & 855 & 1.968 & 1858.02 & 61.91 & 728.39 & 0.085 \\
                     &                       & 955 & 1.719 & 2000.21 & 62.03 & 784.13 & 0.079 \\
                     &                       &1055 & 1.521 & 2137.51 & 62.15 & 837.95 & 0.074 \\
                     &                       &1155 & 1.360 & 2270.53 & 62.25 & 890.10 & 0.070 \\
\hline
\multirow{8}{*}{0.8} & \multirow{8}{*}{0.92} & 455 & 4.692 & 1233.84 & 104.00 & 476.05 & 0.218 \\
                     &                       & 555 & 3.715 & 1408.58 & 104.41 & 543.70 & 0.192 \\
                     &                       & 655 & 3.049 & 1573.07 & 104.72 & 606.93 & 0.173 \\
                     &                       & 755 & 2.569 & 1729.35 & 104.95 & 667.23 & 0.157 \\
                     &                       & 855 & 2.249 & 1885.56 & 105.13 & 720.23 & 0.145 \\
                     &                       & 955 & 1.930 & 2022.65 & 105.30 & 780.39 & 0.135 \\
                     &                       &1055 & 1.708 & 2161.50 & 105.43 & 833.96 & 0.126 \\
                     &                       &1155 & 1.527 & 2161.50 & 105.54 & 885.86 & 0.119 \\
\enddata

\end{deluxetable*}

\section{Disccusion}
\label{5}
Our study presents the implementation of Bondi-Hoyle-Lyttleton accretion in a massive binary system within the \textsc{mesa} stellar evolution framework, where the primary star undergoes a mass loss event, and the companion accretes part of the ejected material through wind accretion. By varying the mass ratio  and orbital period, we quantified how these parameters affect the  accretion rate onto the companion. These results can be placed in the context of both classical and recent work on wind accretion in binaries.

\begin{figure*}
  \includegraphics[trim= 0.0cm 0.0cm 0cm 0.0cm,clip=true,width=1\textwidth]{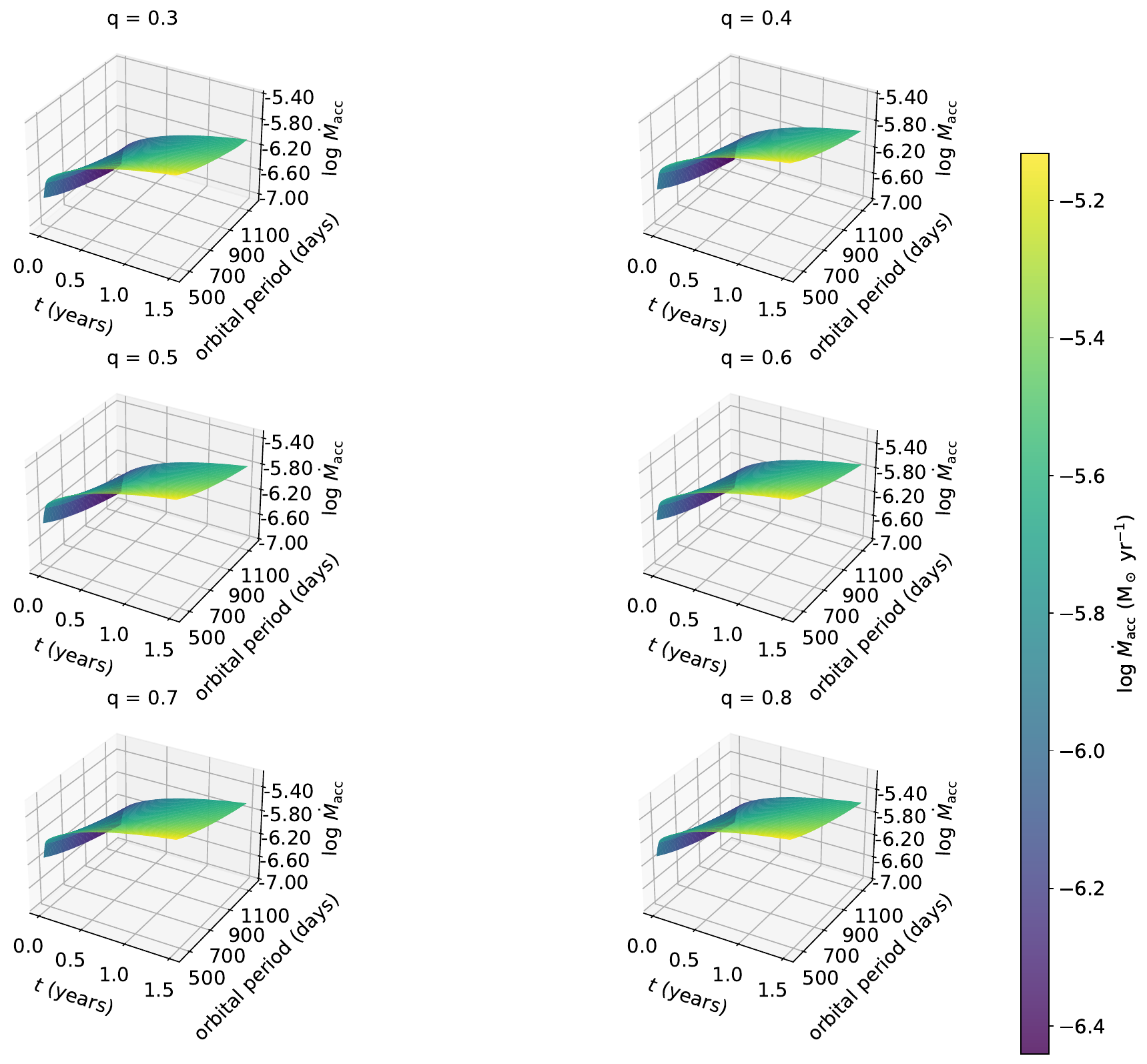} 
  \centering
  \caption{Three-dimensional representation of the wind accretion over a fixed simulation period of 1.5 years. The plot illustrates how the average accretion rate varies as a function of the mass ratio \( q \) and orbital period. Higher mass ratios lead to increased average accretion rates, whereas longer orbital periods result in reduced accretion efficiency. The average accretion rate $\log \dot{M}_{\rm acc}$ is given in $\rm {M_{\odot}}\rm~yr^{-1}$.}
\label{fig_5}
\end{figure*}

Previous hydrodynamical simulations of wind accretion have primarily focused on compact object companions, such as white dwarfs, neutron stars, or black holes in high-mass X-ray binaries (e.g., \citealt{1996ApJ...471..454B}; \citealt{2001ApJ...562..925B}; \citealt{2004A&A...419..335N}; \citealt{2025A&A...695A.117N}). In those systems, the accretor has no intrinsic wind, making them ideal for testing the wind accretion. Our implementation differs by considering a massive main-sequence or evolved star as the accretor, which could potentially develop its own wind, though we also did not include the wind mass loss by the companion star in our simulations. The dependence of average accretion rate on mass ratio \( q \)  found in our simulations is also consistent with trends seen in earlier 3D simulations, such as those by \citet{2022MNRAS.516.3193K}, who demonstrated that accretion is enhanced when the accretor is significantly more massive than the donor. In our models, higher \( q \)   values (i.e., more massive companions) lead to stronger gravitational focusing of the wind material and causes the higher average accretion rates. This supports a general picture where wind accretion becomes more efficient as the companion mass increases, particularly when the wind velocity is not much greater than the orbital velocity.
Furthermore, our results reaffirm the finding by \citet{2020svos.conf..243K} that orbital separation plays a critical role: systems with shorter orbital periods show markedly higher accretion rates, since the accretor resides deeper within the wind acceleration zone and can capture more of the slower, denser material. This was clearly illustrated in our 3D parameter space analysis (Figure~\ref{fig_5}), where lower periods and higher q values produce the most significant average accretion rates.

We base the adopted physical parameters of our simulations on well-studied accreting system $\eta$ Carina \citep[e.g.,][]{2007MNRAS.378.1609K, 2008NewA...13..569K, 2016ApJ...825..105K}. The primary of $\eta$ Carina undergoes extreme mass loss, with inferred rates approaching $\rm \sim 10^{-3}\,M_\odot \,\mathrm{yr^{-1}}$ \citep[e.g.,][]{2010ApJ...723..602K} and an orbital period of $\simeq$ 5.53 years, making it a prime case where wind accretion dominates over RLOF. Motivated by such conditions, we adopted a fiducial mass-loss rate of $10^{-3}\, \rm M_\odot \, \mathrm{yr^{-1}}$ in our model. This value is in the upper range of mass rates proposed for $\eta$ Carina, though during its Great Eruption, rates had been much higher. It provides an extreme yet relatively stable regime that avoids drastic structural changes in the primary. As shown by \citet{2024ApJ...974..124M}, adopting very high mass loss rate value (e.g. $\dot{M} \sim 0.15 \, \rm M_\odot\, \mathrm{yr^{-1}}$) can lead to change in the stellar envelope and significantly alter the primary’s structure, thereby affecting wind accretion efficiency. To prevent such complications, we selected $10^{-3}\, \rm M_\odot\,\mathrm{yr^{-1}}$ as a representative value consistent with high mass-loss episodes. We are  that in massive binaries, after the main-sequence phase, Case A mass transfer tends to dominate in systems with high-mass donors \citep[e.g.,][]{2015A&A...580A..20S, 2015A&A...573A..71K, 2023A&A...672A.198S}. Since our model is entirely based on wind accretion mechanics and motivated by observations from $\eta$~Carina type systems, we aimed to avoid the Case A mass transfer mechanism driven by Roche lobe overflow. To achieve this, we first evolved both stars separately up to the initial post-MS phase, and then implemented them into the binary module in MESA. This setup allows us to isolate and study the role of wind accretion and its effects in massive binary evolution. In future work, we intend to extend our study to a grid of higher mass-loss rates (up to $\sim 1\, \rm M_\odot\,\mathrm{yr^{-1}}$) in order to explore the structural response of the primary and its impact on accretion. The simulation timescale of 1.5 years is chosen to cover the critical interval around periastron passage, where accretion is expected to peak. The adopted companion mass range ($30$–$60 \,\rm  M_\odot$ \citep{2019MNRAS.486..926K})  follows current, though uncertain, estimates for the $\eta$ Carina secondary. Finally, for simplicity, we assumed a circular orbit, as the main goal of this work was to establish a baseline wind accretion model. Future studies should include eccentric orbits, tidal interactions, and magnetic braking, which are expected to have an important impact on the evolution of massive binaries.

\section{Summary}
\label{6}

In this paper, we explored Bondi-Hoyle-Lyttleton accretion in a grid of massive binary systems using \textsc{mesa} simulations, focusing on how the accretion rates depend on stellar parameters, orbital properties, and mass ratio. Our simulations span a range of mass ratios from \( q \) = 0.3 to 0.8, covering systems where the companion star masses range from $M_{\rm 2}$ = 30 to 80 $\rm M_{\odot}$, while keeping the primary star mass fixed $M_{\rm 1}$ = 100 $\rm M_{\odot}$. The results reveal that the accretion rate onto the companion star increases with increasing mass ratio, even though both the mass-loss rate from the primary and the simulation duration (1.5 years) are held constant. This trend is consistent with expectations from wind accretion theory: a more massive companion generates a deeper gravitational potential well, thereby capturing a larger fraction of the wind material ejected by the primary. The enhanced gravitational focusing leads to more efficient accretion as the companion mass increases.

Separately, we observe from Figure~\ref{fig_2} that the accretion rates exhibit a slight upward trend over time, particularly in models with longer orbital periods. This time evolution may be attributed to changes in the orbital configuration during the simulation. Specifically, as the orbital period increases, the orbital velocity of the primary decreases, which can alter the relative velocity between the wind and the companion. This might reduced relative velocity, improve the conditions for gravitational capture, potentially leading to a gradual increase in accretion rate over time. We also investigated how accretion impacts the stellar parameters of the companion star. As the companion gains mass, its luminosity and radius increase, while the effective temperature decreases slightly, indicating that accreted material influences the outer structure of the star. This behavior is further quantified in Figure~\ref{fig_5}, where a 3D surface plot demonstrates how the average accretion rate correlates with both the orbital period and mass ratio. The figure shows that while high \( q \)  values correspond to higher average accretion rates, longer orbital periods tend to suppress accretion efficiency. This trend is physically intuitive, as a larger orbital separation reduces the density and velocity contrast between the wind and the companion's gravitational sphere of influence. Our results also indicate that while the accretion process leads to a decrease in the surface temperature of the companion, it does not induce significant radial expansion. This outcome suggests that the thermal response of the accreting star is moderated by its ability to redistribute the acquired mass without substantial deviation from thermal equilibrium.
This result is in agreement with the companion response to low accretion rates (see \citet{2025ApJ...986..188M}). In contrast, at higher accretion rates, the thermal equilibrium of the accreting star cannot be maintained, resulting in pronounced expansion and structural changes, as shown in \citet{2025ApJ...986..188M}.

Another aspect of our results is the impact of mass loss on the primary star. The reduction in mass leads to a notable decrease in luminosity, a behavior attributed to the expansion of the stellar envelope and subsequent redistribution of internal energy. This response highlights the need to incorporate detailed mass loss prescriptions in stellar evolution models of massive binaries, particularly in cases where high mass-loss rates are involved.

Our derived analytical relation for average accretion rates as a function of orbital period and mass ratio provides a useful tool for estimating wind accretion efficiency in a broader range of systems. The power-law dependence on these parameters emphasizes the strong sensitivity of average accretion rates to the binary configuration, highlighting the importance of considering system-specific factors when modeling MT in binaries.

The outcome of our simulations is the identification of an orbital period threshold $P \lesssim$ 355 days, beyond which RLOF begins to contribute significantly to the MT process alongside wind accretion. In comparison, the existence of such a threshold is consistent with prior studies (e.g., \citealt{2024ApJ...966..103P}). Our models reproduce this transition point, highlighting the role of orbital separation in determining the dominant accretion mechanism. Thus, systems with longer orbital periods rely predominantly on wind accretion, whereas more compact systems transition into a mixed regime where RLOF can dominate. This result aligns with observational studies that identify both wind-fed and RLOF-driven accretion processes in different classes of massive binaries \citet{2024ApJ...966..103P}.

In our above simulations, we do not account for the effects of gravitational wave radiation (which shrinks the orbit), magnetic braking (which removes angular momentum and enhances mass loss), or tidal angular momentum coupling (which synchronizes rotation and affects orbital evolution), all of which can influence the accretion process. Taken together, these findings highlight the complex interplay between binary configuration and accretion efficiency. They have important implications for the evolution of massive binary systems, particularly those that may evolve into high-mass X-ray binaries or progenitors of gravitational wave sources. Future work could explore non-circular orbits, varying wind velocities, and feedback from radiation pressure and spin-up effects, which are all expected to play significant roles in a more realistic wind accretion.

\vspace{0.5cm}
We thank an anonymous referee for very helpful comments.
We acknowledge the Ariel HPC Center at Ariel University for providing computing resources that have contributed to the research results reported in this paper. BM acknowledges support form the AGASS center at Ariel University. The \textsc{mesa} inlists and input files to reproduce our simulations and associated data products are available on Zenodo (10.5281/zenodo.15584880).

\bibliographystyle{mnras}
\bibliography{Ref}{}
\end{document}